\documentclass[aps,prl,preprint,superscriptaddress,showpacs]{revtex4}

\def\be{\begin{equation}}
\def\ee{\end{equation}}
\def\ba{\begin{eqnarray}}
\def\ea{\end{eqnarray}}

\def\nn{\nonumber}
\def\sss{\scriptscriptstyle}

\begin{document}

\preprint{UdeM-GPP-TH-06-143}

\title{\boldmath Looking for $\Delta I = 5/2$ amplitude components in
$B \to \pi \pi$ and $B \to \rho \rho$ experiments}


%
\author{F.\ J.\ Botella}
\affiliation{Departament de F\'{\i}sica Te\`{o}rica and IFIC,
        Universitat de Val\`{e}ncia-CSIC,
        E-46100, Burjassot, Spain}
\author{David London}
\affiliation{Physique des Particules,
        Universit\'{e} de Montr\'{e}al,
	C.P.\ 6128, succ.\ centre-ville,
	Montr\'{e}al, QC, Canada H3C 3J7}
\author{Jo\~{a}o P.\ Silva}
\affiliation{Instituto Superior de Engenharia de Lisboa,
	Rua Conselheiro Em\'{\i}dio Navarro,
	1900 Lisboa, Portugal}
\affiliation{Centro de F\'{\i}sica Te\'{o}rica de Part\'{\i}culas,
	Instituto Superior T\'{e}cnico,
	P-1049-001 Lisboa, Portugal}

\date{\today}

\begin{abstract} 
We discuss how experiments measuring $B \to \pi\pi$ and $B \to \rho
\rho$ may be used to search for a $\Delta I = 5/2$ amplitude
component. This component could be the explanation for a recent
(albeit very tentative) hint from $B ({\bar B}) \to \rho \rho$ decays
that the isospin triangles do not close.
\end{abstract}


\maketitle

Within the standard model (SM), CP violation is due to a complex phase
in the Cabibbo-Kobayashi-Maskawa (CKM) quark mixing matrix. This phase
information can be elegantly encoded in the unitarity triangle
\cite{Wolf,PDG}, in which the interior CP-violating angles are called
$\alpha$, $\beta$ and $\gamma$. Independent measurements of the sides
and angles of the unitarity triangle allow tests of the SM explanation
of CP violation.

The canonical decay mode for measuring $\alpha$ is $B^0(t) \to
\pi^+\pi^-$. However, due to the fact that this decay receives both
tree and penguin contributions, $\alpha$ cannot be extracted cleanly
-- there is penguin ``pollution.'' On the other hand, if one uses
isospin to combine measurements of $B^+ \to \pi^+\pi^0$, $B^0(t) \to
\pi^+\pi^-$ and $B^0(t) \to \pi^0\pi^0$, as well as the CP-conjugate
decays, then the penguin pollution can be removed, and $\alpha$ obtained
cleanly \cite{GL}.

The isospin analysis goes as follows. Due to Bose statistics and the
fact that the final-state pions come from the decay of a spinless
state, they must be in a symmetric isospin configuration. As a result,
the final states are
\ba
\langle \pi^0 \pi^0 | &=& 
\sqrt{\frac{2}{3}} \langle 2, 0 | 
-
\sqrt{\frac{1}{3}} \langle 0, 0 | ~,
\nn\\
\langle \pi^+ \pi^- | &=& 
\sqrt{\frac{1}{3}} \langle 2, 0 |
+
\sqrt{\frac{2}{3}} \langle 0, 0 | ~,
\nn\\
\langle \pi^+ \pi^0 | &=& 
\langle 2, 1 | ~.
\ea
In the SM, short-distance diagrams contribute only to the $\Delta I =
1/2$ and $\Delta I = 3/2$ transitions.  Thus, the physical decay
amplitudes are
\ba
A_{+-} \equiv \langle \pi^+ \pi^- |T| B^0 \rangle
&=&
- \sqrt{\frac{1}{3}} A_{1/2} + \sqrt{\frac{1}{6}} A_{3/2} ~,
\nn\\
A_{00} \equiv \langle \pi^0 \pi^0 |T| B^0 \rangle
&=&
\sqrt{\frac{1}{6}} A_{1/2} + \sqrt{\frac{1}{3}} A_{3/2} ~,
\nn\\
A_{+0} \equiv \langle \pi^+ \pi^0 |T| B^+ \rangle
&=&
\frac{\sqrt{3}}{2} A_{3/2} ~,
\ea
where $A_k$ $(k=1/2, 3/2)$ are the relevant reduced matrix elements.
The parametrization for the CP-conjugate modes is similar, with the
isospin amplitudes replaced by $\bar A_k$. Because there are two
transitions, but three decays, the $B$ decay amplitudes obey a
triangle relation: 
\be
\sqrt{2} A_{+0} = A_{+-} + \sqrt{2} A_{00} ~.
\label{triangle}
\ee
The measurement of the three decays allows one to extract $A_{3/2}$,
while the CP-conjugate decays give ${\bar A}_{3/2}$. However, the
penguin amplitude contributes only to $A_{1/2}$, so that the relative
phase between $A_{3/2}$ and $(q/p){\bar A}_{3/2}$ is $2\alpha$, where
$q/p$ describes $B$--$\bar B$ mixing.  Thus, the penguin pollution has
been removed.

Now, a generic $B\to\pi\pi$ transition contains $\Delta I = 1/2$,
$\Delta I = 3/2$, and $\Delta I = 5/2$ terms, which contribute to the
physical decay amplitudes as
\ba
A_{+-} &=&
- \sqrt{\frac{1}{3}} A_{1/2} + \sqrt{\frac{1}{6}} A_{3/2}
- \sqrt{\frac{1}{6}} A_{5/2} ~,
\nn\\
A_{00} &=&
\sqrt{\frac{1}{6}} A_{1/2} + \sqrt{\frac{1}{3}} A_{3/2}
- \sqrt{\frac{1}{3}} A_{5/2} ~,
\nn\\
A_{+0} &=&
\frac{\sqrt{3}}{2} A_{3/2}
+ \sqrt{\frac{1}{3}} A_{5/2} ~.
\label{full_decomposition}
\ea
The key point is that, in the presence of a nonzero $A_{5/2}$, the
three $B\to\pi\pi$ amplitudes by themselves no longer obey a triangle
relation.  That relation is modified as follows:
\be
\sqrt{2}A_{+0}\left( 1-z\right) =A_{+-}+\sqrt{2}A_{00} ~,
\ee
with
\be
y\equiv \frac{A_{5/2}}{A_{3/2}} = \frac{z}{1+\frac{2}{3}\left(
1-z\right) } ~.
\ee

Although isospin symmetry was mentioned above,
Eqs.~(\ref{full_decomposition}) already take into account any possible
isospin-breaking effects in the decay amplitudes, since the three
isospin amplitudes are enough to encode all the information contained
in the three experimental amplitudes.

Note also that, although $B\to\pi\pi$ decays were described above, the
isospin analysis also holds for each final-state polarization of
$B\to\rho\rho$ decays. In addition, it holds for the decay of any
neutral isospin-1/2 meson. In particular, it applies if the initial
meson is $K$ or $D$.

As noted above, the SM contributes only to the $\Delta I = 1/2$
and $\Delta I = 3/2$ transitions at short distance.  The $\Delta I =
5/2$ transitions arise from rescattering effects, such as the
combination of $A_{1/2}$ with a $\Delta I = 2$ electromagnetic
rescattering of the two pions in the final state.  This is naively
estimated to be of order $|A_{5/2}| \sim \alpha |A_{1/2}|$, where
$\alpha \sim 1/127$ is the electromagnetic coupling constant.  There
are also strong-interaction isospin-violating effects ($m_u \neq
m_d$).

A $\Delta I = 5/2$ contribution was first identified in $K \to \pi\pi$
decays.  In this case, $|A_{1/2}| \sim 20 |A_{3/2}|$ (known as the
$\Delta I = 1/2$ rule), meaning that $|A_{5/2}| \sim 0.1 |A_{3/2}|$,
thus influencing the decay $K^+ \to \pi^+ \pi^0$ \cite{Kpipi}.  A
detailed comparison between theory and experiment is rather involved;
a recent analysis within chiral perturbation theory may be found in
Ref.~\cite{Pich}.

In contrast, in the $B$ system it is expected that $|A_{1/2}| \sim
|A_{3/2}|$ and $A_{5/2}$ is normally discarded (as above, in the
isospin analysis). (Recent analyses including electromagnetic and
strong isospin violation can be found in Ref.~\cite{recent}.) Our main
purpose is to encourage experiments to scrutinize this assumption very
closely, highlighting the fact that current data could be interpreted
as showing some hints of $A_{5/2} \neq 0$.  This is an important issue
since, if $A_{5/2} \ne 0$, the isospin triangles do not close, and the
extraction of $\alpha$ will be affected.

If the SM is valid and the arguments leading to $A_{5/2}=0$ are
correct, then four predictions can be made: 
\begin{enumerate}

\item as noted above, the triangle in Eq.~(\ref{triangle}) and its
conjugate version close.

\item all measurements of $\alpha$ will yield the same result. For
example, the CP phase $\beta$ has already been measured very precisely
in $B^0(t) \to J/\psi K_S$: $\sin 2\beta = 0.726 \pm 0.037$
\cite{sin2beta}, which determines $\beta$ up to a four-fold ambiguity.
The phase $\gamma$ can in principle be cleanly determined through CP
violation in decays such as $B\to DK$ \cite{BDK}, or from a fit to a
variety of other measurements (the latest analysis gives $\gamma =
{58.2^{+6.7}_{-5.4}}^\circ$ \cite{CKMfitter}). The phase $\alpha$ is
then given by $\alpha_{\sss UT} \equiv \pi - \beta - \gamma$. If
$A_{5/2}=0$, then $\alpha_{\rm fit} = \alpha_{\sss UT}$, where
$\alpha_{\rm fit}$ is determined from $B\to\pi\pi$ or $B\to\rho\rho$
decays.

\item the direct CP asymmetry in $B^+ \to \pi^+ \pi^0$ ($C_{+0}$)
vanishes.

\item because there is one more observable than independent parameters
in $B\to\pi\pi$, the interference CP asymmetry parameter in $B^0 \to
\pi^0 \pi^0$ ($S_{00}$), may be written as a function of the other
observables: $F(S_{00}, C_{00}, B_{00}, S_{+-}, C_{+-}, B_{+-},
C_{+0}, B_{+0}) = 0$.  Here $B$, $C$, and $S$ represent the
CP-averaged branching ratio, the direct CP violation and the
interference CP violation, respectively.

\end{enumerate}
Of the four predictions, only the first and fourth are smoking-gun
signals of $A_{5/2} \ne 0$; the others can be violated in the presence
of physics beyond the SM with $A_{5/2} = 0$. The situation is
summarized in Table~\ref{tests}.

\begin{table}[ht]
\caption{Strategies to utilize the experimental observables to
distinguish three cases: neglecting isospin-violations
in the SM (IC-SM);
considering isospin-conserving new physics (NP);
and considering $\Delta I = 5/2$ components.}
\begin{ruledtabular}
\begin{tabular}{lccc}
& IC-SM & NP  &  $\Delta I = 5/2$ \\
\hline
triangle & closes & closes & does not close\\
$\alpha_{\rm fit} - \alpha_{\sss UT}$& $=0$ & $\neq 0$ & $\neq 0$\\
$C_{+0}$ & $=0$ & $\neq 0$ & $\neq 0$\\
$F(S_{00}, \dots)$ & $=0$ & $= 0$ & $\neq 0$\\
\end{tabular}
\end{ruledtabular}
\label{tests}
\end{table}

The most obvious test for a nonzero $A_{5/2}$ is the non-closure of
the isospin triangle. In the following, we examine the present data on
$B({\bar B})\to\pi\pi$ and $B ({\bar B})\to\rho\rho$ decays with this
in mind. In analyzing the $\rho\rho$ data we assume that these
particles are completely longitudinally polarized. This is known
experimentally to be an excellent approximation \cite{f_L=1}. 

Note that, since $A_{5/2}$ is expected to be small, it can only be
seen in those triangles which are relatively flat. This is the case
for the $B ({\bar B})\to\rho\rho$ triangles, since the branching
ratios for $B^0 \to \rho^0 \rho^0$ and ${\bar B}^0 \to \rho^0 \rho^0$
are much less than those of the other decay channels. It is also, by
chance, the case for the $B \to\pi\pi$ triangle, but not for that of
${\bar B} \to\pi\pi$.

\begin{table}[ht]
\caption{Branching ratios $B_f$, direct CP asymmetries $C_f$, and
interference CP asymmetries $S_f$ (if applicable) for the three $B \to
\pi\pi (\rho\rho)$ decay modes. Data comes from
Refs.~\cite{BRs,ACPs,pi+pi-,rho+rho-,rho+rho0,rho0rho0}; averages
(shown) are taken from Ref.~\cite{HFAG}.}
\begin{ruledtabular}
\begin{tabular}{lccc}
& $B_f[10^{-6}]$ & $C_f$  & $S_f$ \\
\hline
$B^+ \to \pi^+ \pi^0$ & $5.5 \pm 0.6$ & $-0.01 \pm 0.06$ & \\
$B^0 \to \pi^+ \pi^-$ & $5.0 \pm 0.4$ & $-0.37 \pm 0.10$ & $-0.50 \pm
    0.12$ \\
$B^0 \to \pi^0 \pi^0$ & $1.45 \pm 0.29$ & $-0.28 \pm 0.40$\\
\hline
$B^+ \to \rho^+ \rho^0$ & $26.4 \pm 6.4$ & $0.09 \pm 0.16$ & \\
$B^0 \to \rho^+ \rho^-$ & $26.2 \pm 3.7$ & $-0.03 \pm 0.17$ & $-0.21 \pm
    0.22$ \\
$B^0 \to \rho^0 \rho^0$ & $\leq 1.1$ & $(-1,1)$ \\
\end{tabular}
\end{ruledtabular}
\label{T:data}
\end{table}

The current $B\to\pi\pi$ and $B\to\rho\rho$ experimental measurements
are shown in Table~\ref{T:data}. This data can be turned into
measurements of the $B \to f$ ($A_f$) and $\bar B \to f$ ($\bar A_f$)
decay amplitudes through:
\ba
|A_f|^2 &\propto& B_f (1+C_f) ~,
\nonumber\\
|\bar A_f|^2 &\propto& B_f (1-C_f) ~.
\ea
The proportionality constants involve two ingredients. First, there is
the phase-space factor $K(m_{\sss B},m_f)$ which is essentially the
same for all amplitudes in each channel. The second factor is the
lifetime of the decaying $B$. Thus, $B_+$ and $B_-$ must be multiplied
by $x = \tau(B^0)/\tau(B^+)$, $1/x = 1.076 \pm 0.008$, due to the
difference between the charged and neutral $B$ lifetimes
\cite{PDG}. We present the norms $|A_f|$ and $|\bar A_f|$ in
Table~\ref{T:fit_isospin} in arbitrary units (i.e.\ we include the
factor $x$ but not $K(m_{\sss B},m_f)$).

\begin{table}[ht]
\caption{\label{T:fit_isospin} The isospin amplitudes in $B ({\bar B})
\to \pi \pi$ and $B ({\bar B}) \to \rho\rho$ (in arbitrary units).}
\begin{ruledtabular}
\begin{tabular}{cccc}
& $\sqrt{2} |A_{+0}|$ & $ |A_{+-}|$ & $\sqrt{2}|A_{00}|$ \\
\hline 
$B \to \pi\pi$: & $3.2 \pm 0.3$ & $1.8 \pm 0.2$ & $1.4 \pm 0.6$ \\
$B \to \rho\rho$: & $7.3\pm 1.5$ & $5.0 \pm 0.8$ & $< 1.5
   \sqrt{1 + C_{00}}$ \\
\hline
& $\sqrt{2} |\bar A_{+0}|$  & $ |\bar A_{+-}|$ & $\sqrt{2}|\bar A_{00}|$ \\
\hline
${\bar B} \to \pi\pi$: & $3.2 \pm 0.3$ & $2.6 \pm 0.2$ & $1.9 \pm 0.5$ \\
${\bar B} \to \rho\rho$: & $6.7 \pm 1.4$ & $5.2 \pm 0.8$ & $< 1.5
  \sqrt{1 - C_{00}}$ \\
\end{tabular}
\end{ruledtabular}
\end{table}

We note in passing that, in addition, for the decays of the neutral
$B$ mesons in which $S_f$ is measured, we also have access to the
relative phase in
\be
\lambda_f \equiv \frac{q}{p} \frac{\bar A_f}{A_f}
= \frac{\pm \sqrt{1 - C_f^2 - S_f^2} + i S_f}{1-C_f} ~,
\ee
where $q/p$ arises from $B$--$\bar B$ mixing. However, we will not use
this information.

In order to see if the isospin triangles close, we proceed as
follows. In the absence of $A_{5/2}$, the triangle relation of
Eq.~(\ref{triangle}) holds. We therefore have
\be
|\sqrt{2} A_{+0}| = | A_{+-} + \sqrt{2} A_{00}| \le  |A_{+-}| +
|\sqrt{2} A_{00}| ~.
\ee
Thus, if $|\sqrt{2} A_{+0}|$ is larger than $|A_{+-}| + |\sqrt{2}
A_{00}|$, the triangle cannot close. The logic is siimilar for the
CP-conjugate triangle.

For the $\pi\pi$ final state we see from the data that the central
values do close both the $B\to\pi\pi$ and ${\bar B}\to\pi\pi$
unitarity triangles (but just barely for $B\to\pi\pi$): $|\sqrt{2}
A_{+0}| = 3.2$, $|A_{+-}| + |\sqrt{2} A_{00}| = 3.2$; $|\sqrt{2} {\bar
A}_{+0}| = 3.2$, $|{\bar A}_{+-}| + |\sqrt{2} {\bar A}_{00}| = 4.5$.

However, the same is not true for $B\to\rho\rho$. Here, the data show
that the $B ({\bar B})\to\rho\rho$ isospin triangles {\it do not}
close (we present a detailed analysis below). This is quite
tantalizing: is it simply a statistical flucturation, or is it a
signal of a $\Delta I=5/2$ component at a level larger than naive
expectations?

Consider $B \to \rho\rho$. The length $\sqrt{2}|A_{00}|$ depends on
the value of $C_{00}$, but for the purposes of illustration, suppose
that $C_{00} = 0$. Then the central values give $|\sqrt{2} A_{+0}| =
7.3$, $|A_{+-}| + |\sqrt{2} A_{00}| < 6.5$, and the triangle does not
close. This situation can be rectified by the inclusion of a $\Delta I
= 5/2$ piece. For various values of $C_{00}$, the data require that
\be
\left\vert y\right\vert =\left\vert \frac{A_{5/2}}{A_{3/2}}\right\vert
\geq \left( 
\begin{array}{ccc}
0.01\pm 0.19 & ; & C_{00}=1 \\ 
0.04\pm 0.19 & ; & C_{00}=0.5 \\ 
0.07\pm 0.19 & ; & C_{00}=0 \\ 
0.11\pm 0.19 & ; & C_{00}=-0.5 \\ 
0.21\pm 0.19 & ; & C_{00}=-1%
\end{array}%
\right) 
\ee
For all values of $C_{00}$, a nonzero $A_{5/2}$ is required by the
central values of the present data. However, a study of the errors
shows that, at present, the effect is not yet statistically
significant -- it is at most at the level of $1\sigma$ ($C_{00} =
-1$).

Turning to ${\bar B} \to \rho\rho$, the present data give
\be
\left\vert \overline{y}\right\vert =\left\vert \frac{\overline{A}_{5/2}}{%
\overline{A}_{3/2}}\right\vert \geq \left( 
\begin{array}{ccc}
0.16\pm 0.21 & ; & C_{00}=1 \\ 
0.06\pm 0.21 & ; & C_{00}=0.5 \\ 
0.01\pm 0.20 & ; & C_{00}=0 \\ 
\text{No Bound} & ; & C_{00}=-0.5 \\ 
\text{No Bound} & ; & C_{00}=-1%
\end{array}%
\right) 
\ee
In this case, a nonzero value of $A_{5/2}$ is required only for
certain values of $C_{00}$ (and the effect is not yet statistically
significant).

This summarizes the present hint for a $\Delta I = 5/2$ piece in $B
\to \rho\rho$ and ${\bar B} \to \rho\rho$ decays, separately.
However, the signals go in opposite directions in each decay: the size
of $A_{5/2}$ in $B \to \rho\rho$ decays increases as $C_{00}$ goes
from $+1$ to $-1$, while ${\bar A}_{5/2}$ in ${\bar B} \to \rho\rho$
decays increases as $C_{00}$ goes from $-1$ to $+1$. As a result, we
may combine information from both sets of data, using
\be
|\sqrt{2} A_{+0}| + |\sqrt{2} {\bar A}_{+0}| \le |A_{+-}| + |{\bar
A}_{+-}| + |\sqrt{2} A_{00}| + |\sqrt{2} {\bar A}_{00}| ~.
\ee
The presence of a $\Delta I = 5/2$ piece is implied if this inequality
is not satisfied. The current data imply that 
\be
y\vee \overline{y} \geq 
\left( 
\begin{array}{ccc}
0.08\pm 0.13 & : & C_{00}=1 \\ 
0.04\pm 0.12 & : & C_{00}=0.5 \\ 
0.04\pm 0.12 & : & C_{00}=0 \\ 
0.04\pm 0.12 & : & C_{00}=-0.5 \\ 
0.08\pm 0.13 & : & C_{00}=-1%
\end{array}%
\right) 
\ee
As above, the present data suggest a nonzero $A_{5/2}$ piece for all
values of $C_{00}$, but the effect is not yet statistically
significant.

In summary, we have shown that if the usual $B ({\bar B}) \to \pi\pi$
or $B ({\bar B}) \to \rho\rho$ isospin triangles do not close, this
may be due to a SM $\Delta I = 5/2$ piece ($A_{5/2}$) at a level much
larger than expected. This is a crucial question since a $A_{5/2}$
piece can also mimic new-physics contributions to other observables,
such as $C_{+0}$ or $\alpha_{\rm fit} - \alpha_{\sss UT}$ (see
Table~\ref{tests}). We have pointed out some strategies to disentangle
$A_{5/2}$ from legitimate new physics.

At present, data on $B ({\bar B})\to\rho\rho$ decays give a hint --
not yet statistically significant -- that the isospin triangles do not
close. The purpose of this letter is to stress the need for
experimental scrutiny of such a signal (and to continue to look for
one in $B ({\bar B}) \to \pi\pi$).  [A probe with $F(S_{00}, \dots)$
is also possible (Table~\ref{tests}), particularly for $B\to\rho\rho$,
and advisable once the data become more precise.]  If this signal
remains, it may be a sign of a SM $\Delta I = 5/2$ piece.

\begin{acknowledgments}
We would like to thank the anonymous referee of \cite{bpipi1} for
making a comment which lead us to think about this problem.  The work
of D.\ L.\ is supported by NSERC of Canada.  F.\ J.\ B.\ is partially
supported by the spanish M.\ E.\ C.\ and FEDER under
FPA2005-01678. J.\ P.\ S.\ is supported in part by the Portuguese
\textit{Funda\c{c}\~{a}o para a Ci\^{e}ncia e a Tecnologia} (FCT)
under the contract CFTP-Plurianual (777), and through the project
POCTI/37449/FNU/2001, approved by the Portuguese FCT and POCTI, and
co-funded by FEDER.
\end{acknowledgments}


\end{document}